# Tetragonal Fe$_2$O: the stable iron oxide at Earth's core conditions

Junjie Jiang [1#], Zhen Zhang [2#], Tongqi Wen [3], Renata M. Wentzcovitch [4-7], Yang Sun [1*]

[1]*Department of Physics, Xiamen University, Xiamen 361005, China*
[2]*Department of Physics and Astronomy, Iowa State University, Ames, Iowa 50011, United States*
[3]*Center for Structural Materials, Department of Mechanical Engineering, The University of Hong Kong, China*
[4]*Department of Applied Physics and Applied Mathematics, Columbia University, New York, NY 10027, USA*
[5]*Department of Earth and Environmental Sciences, Columbia University, New York, NY 10027, USA*
[6]*Lamont–Doherty Earth Observatory, Columbia University, Palisades, NY 10964, USA*
[7]*Data Science Institute, Columbia University, New York, NY 10027, USA*
(Dated: November 18, 2025)

The Fe-O system is fundamental to understanding the composition and properties of the Earth's core. Recent studies have suggested the possible existence of stable, iron-rich Fe$_n$O compounds at around 215 GPa. Here, we performed crystal-structure searches and fully anharmonic free-energy calculations to investigate the Fe-FeO system under inner-core conditions. We identified Fe$_2$O as a stable phase and constructed its high P-T phase diagram. Fe$_2$O undergoes a hexagonal-to-tetragonal transition with increasing pressure and temperature. It remains thermodynamically stable against decomposition into Fe and FeO from 200 to 400 GPa and at high temperatures. Although oxygen has been considered nearly absent in the inner core due to its limited solubility, these results suggest that oxygen can, in fact, be incorporated into the solid inner core in the form of an Fe+Fe$_2$O mixture, and can match PREM densities for 53 mol% Fe$_2$O. Our work has the potential to lead to a significant revision of the current understanding of the core's structure and composition.

**I. Introduction**

The Fe-O system plays a vital role in understanding the composition and properties of the Earth's core, as iron is its major constituent and oxygen is a plausible light element candidate. High-pressure melting experiments have demonstrated that eutectic Fe-FeO liquids can dissolve up to ~15 wt% oxygen under inner core boundary (ICB) conditions [1]. Modeling of core–mantle differentiation also suggests that oxygen is an essential light element in the liquid outer core (OC) [2]. Under oxygen-rich conditions, the Fe-O system exhibits rich stoichiometry, such as the FeO$_2$ under lower-mantle conditions [3–5]. However, oxygen has traditionally been regarded as nearly absent in the inner core (IC) [6], due to its limited solubility in solid iron [7,8] and the absence of experimentally observed iron-rich Fe-O compounds even at core pressure [9,10].

This long-standing view has been challenged by the recent discoveries of Fe-rich Fe-O compounds under core pressures. Liu et al. [11] performed ab initio crystal structure prediction (CSP) and identified several stable iron-rich Fe-O compounds with hexagonal lattices at 215 GPa and 0 K. Among them, a hexagonal phase in a supercell of Fe$_{28}$O$_{14}$ shows excellent agreement with the X-ray diffraction patterns of quenched experimental samples [11]. The stability of hexagonal phases was confirmed by Jang et al. [12] with a few ab initio techniques. Weerasinghe et al. [13] predicted several iron-rich Fe-O compounds at 100-500 GPa and 0 K. Extending the CSP to the 500 GPa to 3 TPa pressure range revealed the the tetragonal I4/mmm Fe$_2$O phase remains stable throughout this entire pressure range [14]. Although finite-temperature effects were not considered in previous CSP, these findings reveal the possible stability and structural diversity of iron-rich Fe-O phases under core conditions.

From a theoretical perspective, the Fe–O system presents a persistent challenge due to the localized and strongly correlated 3d electrons of iron. It was demonstrated that the LDA+U$_{sc}$ method can accurately describe the insulating B1 and iB8 phases of FeO, while calculations based on Mermin functional are required to incorporate the electronic entropy in the metallic nB8 phase near core-mantle boundary pressures [15]. Zhang et al. [16] successfully determined the B2−B8 phase boundary of FeO consistent with experimental data at core conditions using the purely PBE-GGA functional, which effectively captures both electronic and anharmonic vibrational properties in metallic, nonmagnetic FeO at high pressures and temperatures. It has been shown that the PBE-GGA functional can also accurately describe the melting curve and equation of state of pure Fe under Earth's core conditions [17–19]. Collectively, these studies establish a robust ab initio framework for investigating the Fe-FeO system under Earth's core conditions.

---

[#]Equal contribution.
[*]Email: yangsun@xmu.edu.cn

In this work, we study stable Fe-rich FeO phases at 350 GPa using CSP with adaptive genetic algorithm (AGA). We investigate phase transition using the fully anharmonic free energy calculation over a wide P-T range (200-400 GPa and 0-4000 K) relevant to the Earth's core. The thermodynamic stability with respect to decomposition into Fe and FeO was systematically investigated.

## II. Methods
### A. Crystal structure search

The crystal structural search for Fe-rich $Fe_nO$ compounds was performed using the AGA [20–22] at 350 GPa. The original AGA combines *ab initio* calculations with auxiliary interatomic potentials described by the embedded-atom method in an adaptive manner to ensure both efficiency and accuracy. The searches were constrained only by the chemical composition, without any prior assumption regarding the Bravais lattice type, symmetry, atomic basis, or unit-cell dimensions. A broad range of iron-rich compositions $Fe_xO_y$ (x:y = 6:5, 5:4, 4:3, 3:2, 5:3, 2:1, 5:2, 3:1, 4:1, 5:1, 6:1, 7:1, 8:1, 9:1) were explored, with up to 25 atoms per unit cell.

Each AGA run consisted of a genetic algorithm (GA) loop accelerated by the interatomic potential and a density functional theory (DFT) loop used to iteratively refine that potential. The GA search maintained a candidate pool of 64 structures. In each generation, 16 offspring were generated from the parent pool via the mating operation, and the pool was updated by retaining the 64 lowest-energy structures.

For a given auxiliary potential, the GA search was carried out for 1000 consecutive generations. After each GA loop, 16 structures were randomly selected for static DFT calculations to refine the potential. This adaptive process was repeated 80 times, after which all distinct structures were further optimized using full DFT relaxations.

### B. First-principles calculations

All first-principles calculations were performed using the projector augmented-wave (PAW) method as implemented in the VASP code [23–25]. The exchange-correlation energy was treated using the generalized gradient approximation (GGA) with the Perdew-Burke-Ernzerhof (PBE) functional [26]. As demonstrated in previous work [16], the PBE-GGA functional accurately predicts the nB8-B2 phase boundary in metallic FeO at core conditions, validating the applicability of standard DFT for this endmember in the metallic regime and providing a framework that can be extended to related Fe-FeO phases, e.g., $Fe_2O$. The PAW potentials have $3d^74s^1$ and $2s^22p^4$ valence configurations for Fe and O atoms, respectively. A plane-wave kinetic energy cutoff of 400 eV was employed. Finite electronic temperature effects were incorporated using the Mermin functional [27,28].

For static calculations, the Brillouin zone was sampled using a Γ-centered k-point mesh with a reciprocal-space resolution of $2\pi \times 0.033$ Å$^{-1}$. The self-consistent field (SCF) energy convergence criterion was set to $10^{-8}$ eV, and ionic relaxations were performed until the residual forces on every atom were below 0.001 eV/Å.

### C. Phonon calculations

Harmonic phonons were computed using the finite-displacement method implemented in Phonopy [29,30]. Anharmonic phonons were computed using the phonon quasiparticle (PHQ) approach [31,32], as implemented in the phq [33] package.

The PHQ approach models a system with fully interacting phonons as an effective ensemble of noninteracting phonon quasiparticles (PHQs) [17,34], each characterized by a renormalized frequency $\widetilde{\omega}_{qs}$ and a linewidth $\Gamma_{qs}$. Numerically, a PHQ is obtained from the mode-projected velocity autocorrelation function (VAF) [31,32],

$$\langle V_{qs}(0)V_{qs}(t) \rangle = \lim_{\tau \to \infty} \frac{1}{\tau} \int_0^\tau V_{qs}^*(t')V_{qs}(t'+t)dt', \quad (1)$$

where $V_{qs}(t) = \sum_{i=1}^N \sqrt{M_i}\, \bm{v}_i(t) e^{i\bm{q}\cdot \bm{R}_i} \cdot \hat{\bm{e}}_{qs}^i$ is the mass-weighted and $qs$-mode-projected velocity. Here $M_i$, $\bm{R}_i$, and $\bm{v}_i$ ($i = 1, ..., N$) denote the atomic mass, coordinate, and velocity from *ab initio* MD (AIMD) trajectories of an N-atom supercell, respectively; $\hat{\bm{e}}_{qs}^i$ is the harmonic phonon eigenvector. For a well-defined PHQ, the VAF follows an exponentially damped cosine behavior, $A_{qs}\cos(\widetilde{\omega}_{qs}t)e^{-\Gamma_{qs}t}$, where $A_{qs}$ is the initial oscillation amplitude.

AIMD simulations for anharmonic phonons were performed in the NVT ensemble controlled by Nosé thermostat [35,36], with a time step of 1 *fs*. The electronic temperature ($T_{el}$) was set to be the same as the ionic temperature (T) using the Mermin functional [27,28]. Simulations for hexagonal $Fe_2O$ and ε-Fe were run for 1 *ps*, and those for tetragonal $Fe_2O$ for 0.5 *ps*. It was verified that a shorter trajectory length is sufficient for the tetragonal phase to achieve convergence of the vibrational entropy and quasiparticle properties.

The same supercells were used for harmonic and anharmonic phonon calculations. A 6×6×2 supercell containing 216 atoms was used for hexagonal $Fe_2O$, and a 6×6×2 supercell containing 144 atoms for ε-Fe, with *k*-mesh in both structures sampled at the Γ-point. For tetragonal $Fe_2O$, a 4×4×4 supercell containing 192 atoms and a 1×1×2 Monkhorst-Pack *k*-mesh [37] was employed. These parameters were tested to be sufficient to ensure convergence of the vibrational frequencies. Phonon spectra were subsequently interpolated into a 30×30×30 $\boldsymbol{q}$-point grid to approach the thermodynamic limit.

### D. Anharmonic properties

With anharmonic phonon spectra, the vibrational entropy was computed within the phonon gas model (PGM) [31,32,38]:

$$S_{\mathrm{vib}}(T) = k_B \sum_{\boldsymbol{q}s}[(n_{\boldsymbol{q}s} + 1)\ln(n_{\boldsymbol{q}s} + 1) - n_{\boldsymbol{q}s}\ln n_{\boldsymbol{q}s}], \quad (2)$$

where $n_{\boldsymbol{q}s} = \left[\exp\left(\frac{\hbar\widetilde{\omega}_{\boldsymbol{q}s}(T)}{k_B T}\right) - 1\right]^{-1}$. The temperature dependent phonon frequencies $\widetilde{\omega}_{\boldsymbol{q}s}(T)$ at constant volume were obtained by interpolating the computed harmonic phonon and renormalized phonon frequencies at several AIMD-sampled temperatures, to the quadratic polynomial function of T [16,32,39].

The Helmholtz free energy at constant volume, including contributions from both electronic and vibrational entropy, is expressed as [16,40]

$$F(T) = E_0 + \frac{1}{2}\sum_{\boldsymbol{q}s}\hbar\omega_{\boldsymbol{q}s} - \int_0^T [S_{el}(T') + S_{vib}(T')]dT', \quad (3)$$

where $E_0$ is the static energy, and $\frac{1}{2}\sum_{\boldsymbol{q}s}\hbar\omega_{\boldsymbol{q}s}$ is the zero-point energy obtained from the harmonic phonon frequencies at zero temperature. The electronic entropy $S_{el}(T)$ was computed from the Mermin functional at several electronic temperatures and subsequently fitted to a second-order polynomial in T [16,39]. For temperature-invariant harmonic phonon frequencies, Eq. (3) becomes equivalent to the standard quasi-harmonic free energy formula.

## III. Results
### A. Convex hull of the Fe-FeO system

To determine the stable phases of Fe-FeO system under Earth's core conditions, we first performed the AGA crystal structure search at 350 GPa. ε-Fe and nB8-FeO are used as the endmembers of the Fe-FeO system. The relative stability of $Fe_xO_y$ compounds was evaluated from the formation enthalpy, defined as

$$\Delta H_f = \frac{H(Fe_xO_y) - yH(FeO) - (x-y)H(Fe)}{x+y}, \quad (4)$$

where $H$ is the enthalpy of a given structure, and $x$ and $y$ are the mol fractions of Fe and O, respectively.

The convex hull at 350 GPa differs markedly from that at 215 GP reported in [11], as shown in Fig. 1(a). At 215 GPa, the stable $Fe_3O$, $Fe_2O$, $Fe_3O_2$ and $Fe_4O_3$ phases all adopt hexagonal lattices, with hexagonal $Fe_2O$ having the lowest formation energy. At 350 GPa, only two phases, $Fe_2O$ and $Fe_6O_5$, remain stable, with most Fe-rich $Fe_2O$ becoming a tetragonal phase. The emergence of the tetragonal phase significantly lowers the formation enthalpy on the convex hull, causing the hexagonal phases stable at 215 GPa to become metastable at 350 GPa. In addition, tetragonal lattices are not limited to the $Fe_2O$ composition. With compositions of $Fe_5O_2$ and $Fe_5O_3$, the tetragonal structure also becomes more stable than the hexagonal structure, indicating a systematic stabilization of the tetragonal phase at 350 GPa. This particular tetragonal I4/$mmm$ $Fe_2O$ structure was also shown to be stable at 0 K in the range of 0.5 to 3 TPa [14].

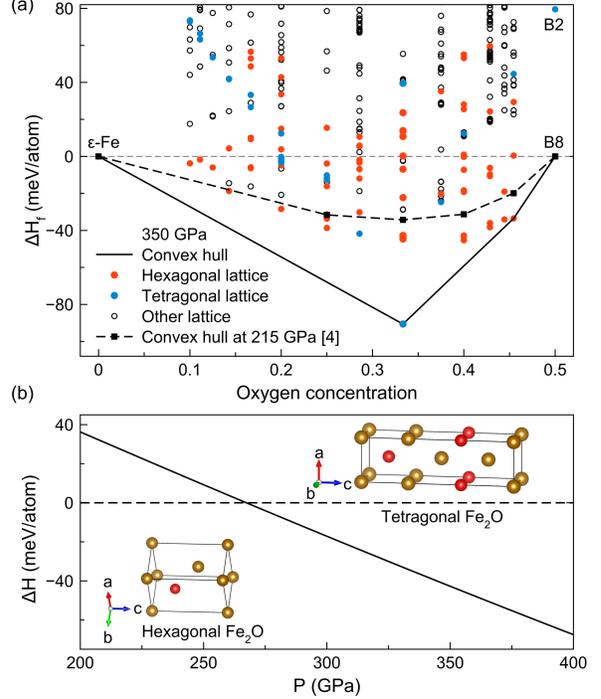

**Figure 1.** (a) Formation enthalpies of Fe-O compounds referenced to ε-Fe and B8-FeO. Colored markers denote crystal structures at 350 GPa: red filled circles for hexagonal lattices, blue filled circles for tetragonal lattices, and black open circles for other lattices. The solid line corresponds to the convex hulls at 350 GPa. The dashed line with black squares indicates the convex hull at 215 GPa [11]. (b) Enthalpy difference between hexagonal (P$\bar{3}$m1) and tetragonal I4/$mmm$ $Fe_2O$ phase, i.e., $\Delta H = H^{Tet} - H^{Hex}$. Negative values indicate that the tetragonal phase is more favorable, while positive values indicate the opposite. Insets show the crystal structures of the two phases, with Fe atoms in brown and O atoms in red.

The marked change in the shape of convex hull indicates a pressure-induced structural transition between the hexagonal and tetragonal phases of $Fe_2O$. Fig. 1(b) shows the enthalpy difference as a function of pressure. The hexagonal phase transforms into the tetragonal one at higher pressure, with a transition point at 263 GPa. The variation of $\Delta H$ between hexagonal and tetragonal phases under compression is dominated by the $\Delta(PV)$ term (Table S1). By fitting the E-V curves to obtain the equation of states, we find the bulk modulus of the tetragonal phase systematically smaller than that of the hexagonal phase, indicating the tetragonal phase is more compressible and thus becomes energetically favored at 350 GPa through the PV contribution (Fig. S1a-b).

The crystal structures of hexagonal and tetragonal phases are shown in the insets of Fig. 1(b). Both phases consist of single-species (Fe-only or O-only) layers stacked along the z-axis, with O-only layers always separated by Fe-only ones. The sixfold-coordinated hexagonal $Fe_2O$ ($P\bar{3}m1$) has a primitive cell containing 3 atoms, while the eightfold-coordinated tetragonal $Fe_2O$ ($I4/mmm$) possesses a conventional unit cell with 6 atoms. The stacking sequence of atomic layers in the hexagonal phase ($A^{Fe}B^{O}C^{Fe}$) differs from that in the tetragonal lattice ($A^{Fe}B^{O}A^{Fe}B^{Fe}A^{O}B^{Fe}$). Each Fe is coordinated by three O atoms in the hexagonal phase, increasing to four in the tetragonal phase. Both phases are non-magnetic and metallic as shown in Fig. S1(c).

### B. Phase transition of $Fe_2O$

To further assess the temperature effect on the phase transition, we performed fully anharmonic free energy calculations. Harmonic phonon dispersions of the hexagonal and tetragonal phases were first computed at six volumes, covering the 200-400 GPa pressure range. Although all volumes investigated yield dynamically stable harmonic phonons, conventional quasi-harmonic free energies can be insufficient to accurately determine the phase boundary due to their neglected lattice anharmonic effects, which become pronounced at high temperatures. Therefore, the PHQ+PGM approach [32] was employed to obtain the full anharmonic free energy.

Anharmonic phonon dispersions in both hexagonal and tetragonal $Fe_2O$ are well-defined; thus, the corresponding VAFs computed using Eq. (1) can be well described by exponentially decaying cosine functions (Fig. 2(c-d)). By interpolating the q-mesh based on PHQ results, anharmonic phonon spectra are obtained as shown in Fig. 2(a-b). Significant differences between harmonic and anharmonic phonon dispersions are observed for both hexagonal and tetragonal phases. In particular, anharmonic effects induce mode-dependent frequency shifts: acoustic branches generally soften ($\Delta\omega_{q,s} < 0$) while optical branches generally harden ($\Delta\omega_{q,s} > 0$). Modes near the Γ point are only slightly shifted ($\Delta\omega_{q,s} \approx 0$).

AIMD simulations are performed over a wide range of V-Ts, covering the corresponding P-T conditions indicated in Fig. S2. In the considered P-T range, all harmonic and anharmonic phonons are stable, with no imaginary frequencies observed, as shown in Fig. S3 and S4 for hexagonal and tetragonal $Fe_2O$, respectively. The dispersions exhibit a continuous and systematic evolution across modes, generally reflecting the progression from 0 K harmonic phonons to anharmonic phonons at temperatures up to 4000 K.

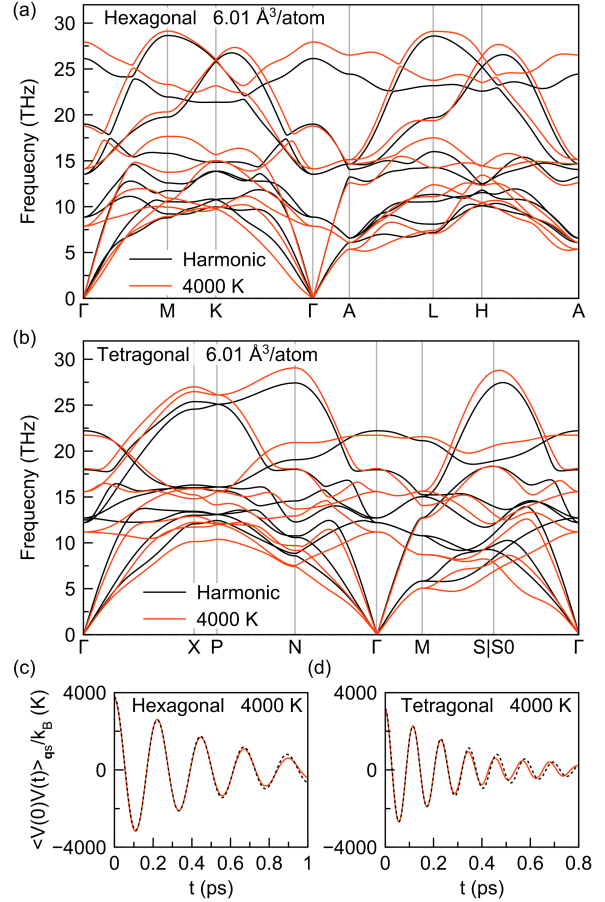

**Figure 2.** Harmonic (black lines) and anharmonic (red lines) phonon dispersions at 4000K of (a) hexagonal and (b) tetragonal $Fe_2O$ at a volume of 6.01 Å³/atom. Mode-projected VAFs for (c) hexagonal and (d) tetragonal $Fe_2O$ calculated based on AIMD simulations (red solid lines) and fitted to decaying cosine functions (black dotted lines).

The vibrational entropy of $Fe_2O$ is obtained in the thermodynamic limit within the phonon gas model using Eq. (2). The electronic entropy is also computed due to the metallic nature of $Fe_2O$ system. The total Helmholtz free energy at each volume was then obtained via the entropy integration method (EIM) expressed in Eq. (3) [12,34]. As shown in Fig. 3(a) and 3(b), the electronic entropy, $S_{el}(V,T)$, increases almost linearly with temperature, while the vibrational entropy, $S_{vib}(V,T)$, exhibits nonlinear behavior, with the latter being dominant in magnitude. Both entropic contributions decrease was the volume decreases At the same V-T conditions, the tetragonal phase consistently exhibits larger $S_{el}(V,T)$ and $S_{vib}(V,T)$ than the hexagonal phase, indicating stronger entropic effects. These enhanced entropic contributions accumulate upon integration (the third term in Eq. (3)), resulting in increased stability of the tetragonal phase at higher temperatures. As shown in Fig. 3(c), the $F(V,T)$ curves intersect at 0 K and 1000 K, where the hexagonal Fe$_2$O exhibits lower free energy at larger volumes. However, as the temperature rises to 2000-4000 K, the curves separate, and the tetragonal Fe$_2$O exhibits lower free energy across all volumes.

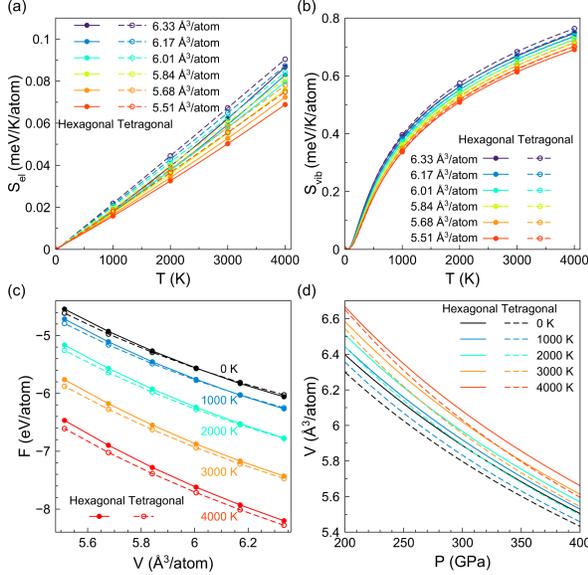

**Figure 3.** (a) Electronic entropy $S_{el}(V,T)$ and (b) vibrational entropy $S_{vib}(V,T)$ as functions of temperature at different volumes. (c) Helmholtz free energy $F(V,T)$ as functions of volume at different temperatures. (d) Equation of state at different temperatures. Solid circles and solid lines correspond to the hexagonal Fe$_2$O phase, while open circles and dashed lines correspond to the tetragonal phase. All circles indicate the computed V-T conditions.

The Helmholtz free energies as a function of volume at each temperature were fitted using the third-order Birch-Murnaghan isothermal equation of state. The pressure was then derived from the fitted curves according to $P = -\left(\frac{\partial F}{\partial V}\right)_T$. As shown in Fig. 3(d), the resulting EOS demonstrates that the tetragonal phase possesses smaller volumes than the hexagonal phase under the same P-T conditions. The Gibbs free energy was subsequently obtained as

$$G(P,T) = F(V,T) + P(V,T)V, \quad (5)$$

by combining the fitted EOS with the Helmholtz free energies at corresponding volumes. The Gibbs free energy difference is shown in Fig. 4. With increasing compression, $\Delta G^{Tet-Hex}(P,T)$ decreases, indicating enhanced relative stability of the tetragonal phase at elevated P-T conditions. The hexagonal − tetragonal phase boundary of Fe$_2$O is shown in Fig. 4. The calculated boundary exhibits a negative Clapeyron slope, $\frac{dP}{dT} = \frac{\Delta S}{\Delta V}$, consistent with relatively larger entropy and smaller volume of the tetragonal phase at higher temperature.

The negative Clapeyron slope of this Fe$_2$O phase boundary resembles that of FeO [16]. This similarity can be rationalized by the structural similarity between the two systems: the low-temperature hexagonal phase of Fe$_2$O shares the same lattice system as B8-FeO, while the high-temperature tetragonal phase is body-centered with a similar motif to B2-FeO. Such correspondence is reminiscent of the hcp−bcc transition in many elemental metals [41,34,42,43].

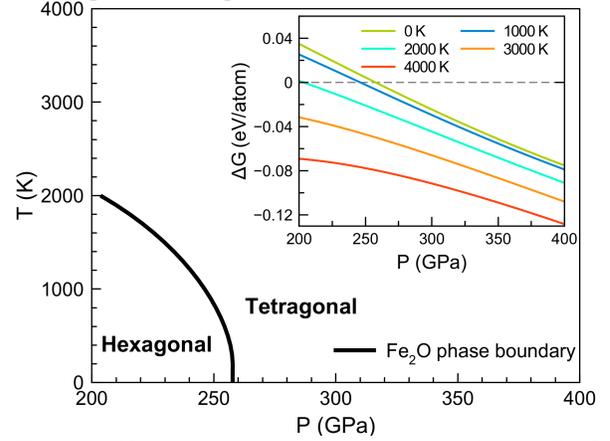

**Figure 4.** Hexagonal-tetragonal phase boundary of Fe$_2$O. The inset shows the Gibbs free energy difference between the hexagonal and tetragonal Fe$_2$O phases ($\Delta G = G^{Tet} - G^{Hex}$) as a function of pressure at various temperatures. Negative values indicate that the tetragonal phase is more stable, while positive values indicate the opposite.

To quantify the anharmonic effect in this system, we compare the Gibbs free energy difference between the quasi-harmonic approximation ($\Delta G^{QHA}$) and the fully anharmonic PHQ approach ($\Delta G$) in Fig. S6. The only distinction between the two approaches lies in the phonon frequencies used to compute the vibrational entropy in Eq. (2), where $\Delta G^{QHA}$ employs temperature-independent harmonic frequencies, whereas $\Delta G$ uses the temperature-dependent renormalized frequencies. At each temperature, $\Delta G$ consistently lies below $\Delta G^{QHA}$, indicating that anharmonicity further stabilizes the tetragonal phase and broadens its stability field. The separation between $\Delta G$ and $\Delta G^{QHA}$ increases with temperature, highlighting the enhancement of anharmonic effects at high temperatures. At 4000 K, the free energy difference from the two approaches can be more than 50 meV/atom. For reference, the free energy difference of bcc and hcp Fe is ~10s meV/atom under IC conditions [19,44,45]. These highlight the critical importance of incorporating anharmonic contributions for an accurate description of the high-temperature phase behavior in Earth's core.

### C. Thermal stability

Combing the current and previous CSP [11,14,46] work, the $Fe_2O$ represents a stable composition with respect to the decomposition into Fe and FeO from 215 GPa to 3 TPa at 0 K. A critical criterion of the thermodynamic stability of $Fe_2O$ is its ability to resist decomposition under high-temperature conditions relevant to IC. Thus, we evaluated its stability with respect to decomposition into $\varepsilon$-Fe and the FeO, which are the nearest stable phases under IC conditions. The Gibbs free energy of $\varepsilon$-Fe was calculated at the P-T conditions shown in Fig. S2, with anharmonic phonon spectra presented in Fig. S5. FeO undergoes a B8−B2 phase transition under core conditions, as reported in [10]. Accordingly, we calculated the Gibbs free energy change of the decomposition reactions $\Delta G_{decomp}$ under various P-T conditions, considering the relevant stable phases of FeO.

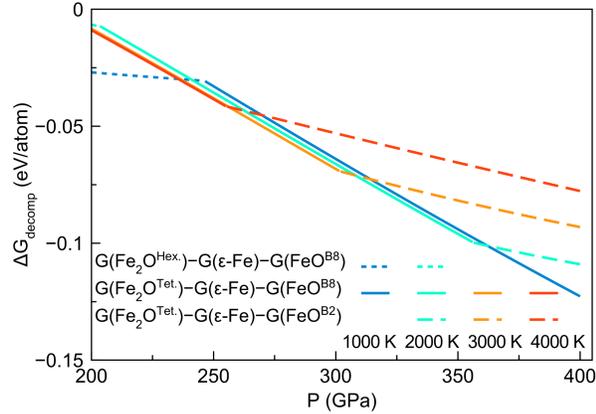

**Figure 5.** Anharmonic Gibbs free energy change for the decomposition reactions of $Fe_2O$ at various temperatures. Negative values indicate that $Fe_2O$ is stable relative to its decomposition products. The reactions considered are: (1) $Fe_2O^{Hex} \leftrightarrow FeO^{B8} + \varepsilon\text{-Fe}$ (dotted line), (2) $Fe_2O^{Tet} \leftrightarrow FeO^{B8} + \varepsilon\text{-Fe}$ (solid line), and (3) $Fe_2O^{Tet} \leftrightarrow FeO^{B2} + \varepsilon\text{-Fe}$ (dashed line).

As shown in Fig. 5, $\Delta G_{decomp}(P,T)$ remains negative across the entire investigated P-T range, indicating that $Fe_2O$ is thermodynamically stable and does not decompose under IC conditions. This is consistent with recent experimental discoveries of iron-rich $Fe_nO$ compounds formed from the reaction between Fe and $FeO/Fe_2O_3$ at ~215 GPa [11]. According to the current phase diagram, the complex $Fe_{28}O_{14}$ phase synthesized by ultrafast laser-heating process and quenched to room temperature [11] may represent an intermediate state, corresponding to a more disordered stacking configuration during the formation of the stable $Fe_2O$ phase. The hexagonal structure of quenched $Fe_{28}O_{14}$ resembles the stable hexagonal phase at low temperatures.

## IV. Discussion

Oxygen has been proposed as a likely light element alloyed with Fe in the liquid OC [2], but is often considered to be nearly absent in the solid IC [6]. This view mainly arises from the lack of experimentally or theoretically identified Fe-rich Fe–O compounds that remain stable under IC conditions. Our results reveal that $Fe_2O$ is thermodynamically stable against decomposition throughout the entire range of IC conditions. Therefore, oxygen can be incorporated into the solid IC in the form of stoichiometric $Fe_2O$. Such incorporation challenges the long-standing view that oxygen resides exclusively in the liquid OC and suggests that the oxygen distribution, as well as other light elements between the OC and IC should be re-evaluated.

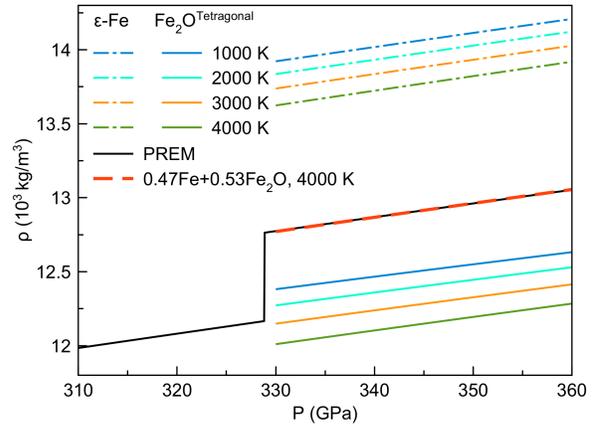

**Figure 6.** Calculated densities of tetragonal $Fe_2O$ (solid lines) and $\varepsilon$-Fe (dash-dot lines) at various temperatures. The densities from the PREM for the IC (dotted line) and OC (dashed line) are shown for comparison.

Figure 6 shows the densities of tetragonal $Fe_2O$ and $\varepsilon$-Fe obtained from EOS fitted from the anharmonic F-V curves under IC conditions. The density of pure iron is higher than that of the IC, implying the presence of light elements—a well-established characteristic of Earth's core structure [47]. The density of tetragonal $Fe_2O$ is lower than that of the IC. A mixture of Fe and $Fe_2O$ in approximately equal proportions can reproduce the observed IC density, with the slope also aligning well with the PREM data. This compositional model, i.e., $FeO_x$, with $x$~0.35, is exceedingly rich in O compared to previously accepted ones, but represents a plausible compositional model for the solid IC when evaluated solely on the basis of density constraints.

## V. Conclusion

In summary, AGA crystal structure searches confirm $Fe_2O$ as the stable phase on the Fe–FeO convex hull with highest Fe composition at 350 GPa. $Fe_2O$ undergoes a structural transition from a hexagonal to a tetragonal phase with increasing pressure from 215 GPa to 350 GPa. Free-energy calculations accounting for full anharmonicity clarify the hexagonal-tetragonal phase boundary in $Fe_2O$, revealing a negative Clapeyron slope. These calculations also show that $Fe_2O$ is thermodynamically stable against decomposition into Fe and FeO under IC conditions over 200–400 GPa and 0–4000 K. The prediction of stable $Fe_2O$ implies that O can be incorporated into the solid inner core in the form of an $Fe+Fe_2O$ mixture, whose density for average composition $FeO_{0.346}$ matches well PREM inner core densities. The existence of $Fe_2O$ would substantially modify current models of the inner core and suggests oxygen plays a critical role that has likely been overlooked in the understanding of the core.

**Acknowledgements**
Work at Xiamen University was supported by the National Natural Science Foundation of China (Grants T2422016 and 42374108). The work at Columbia University was supported by the Gordon and Betty Moore Foundation Award GBMF12801 (doi.org/10.37807/GBMF12801) and NSF Grants (EAR-2000850 and EAR-1918126). Shaorong Fang and Tianfu Wu from the Information and Network Center of Xiamen University are acknowledged for their help with Graphics Processing Unit computing. Some *ab initio* simulations were performed on Bridges-2 system at PSC, the Anvil system at Purdue University, the Expanse system at SDSC, and the Delta system at NCSA through allocation DMR180081 from the Advanced Cyberinfrastructure Coordination Ecosystem: Services & Support (ACCESS) program, which is supported by NSF Grants No. 2138259, No. 2138286, No. 2138307, No. 2137603, and No. 2138296. The supercomputing time were also supported by the Opening Project of the Joint Laboratory for Planetary Science and Supercomputing, Research Center for Planetary Science, and the National Supercomputing Center in Chengdu (Grants No. CSYYGS-QT-2024-15).

## References

[1] K. Oka, K. Hirose, S. Tagawa, Y. Kidokoro, Y. Nakajima, Y. Kuwayama, G. Morard, N. Coudurier, and G. Fiquet, Melting in the Fe-FeO system to 204 GPa: Implications for oxygen in Earth's core, American Mineralogist **104**, 1603 (2019).

[2] J. Badro, J. P. Brodholt, H. Piet, J. Siebert, and F. J. Ryerson, Core formation and core composition from coupled geochemical and geophysical constraints, Proc. Natl. Acad. Sci. U.S.A. **112**, 12310 (2015).

[3] Q. Hu, D. Y. Kim, W. Yang, L. Yang, Y. Meng, L. Zhang, and H.-K. Mao, FeO2 and FeOOH under deep lower-mantle conditions and Earth's oxygen–hydrogen cycles, Nature **534**, 241 (2016).

[4] M. Nishi, Y. Kuwayama, J. Tsuchiya, and T. Tsuchiya, The pyrite-type high-pressure form of FeOOH, Nature **547**, 205 (2017).

[5] J. Liu, Q. Hu, W. Bi, L. Yang, Y. Xiao, P. Chow, Y. Meng, V. B. Prakapenka, H.-K. Mao, and W. L. Mao, Altered chemistry of oxygen and iron under deep Earth conditions, Nat Commun **10**, 153 (2019).

[6] K. Hirose, B. Wood, and L. Vočadlo, Light elements in the Earth's core, Nat Rev Earth Environ **2**, 645 (2021).

[7] D. Alfè, M. J. Gillan, and G. D. Price, Constraints on the composition of the Earth's core from ab initio calculations, Nature **405**, 172 (2000).

[8] Z. Wu, C. Gao, F. Zhang, S. Wu, K.-M. Ho, R. M. Wentzcovitch, and Y. Sun, Ab initio superionic-liquid phase diagram of $Fe_{1-x}O_x$ under Earth's inner core conditions, arXiv **2410.23557** (2024).

[9] H. Ozawa, K. Hirose, S. Tateno, N. Sata, and Y. Ohishi, Phase transition boundary between B1 and B8 structures of FeO up to 210GPa, Physics of the Earth and Planetary Interiors **179**, 157 (2010).

[10] H. Ozawa, F. Takahashi, K. Hirose, Y. Ohishi, and N. Hirao, Phase Transition of FeO and Stratification in Earth's Outer Core, Science **334**, 792 (2011).

[11] J. Liu, Y. Sun, C. Lv, F. Zhang, S. Fu, V. B. Prakapenka, C. Wang, K. Ho, J. Lin, and R. M. Wentzcovitch, Iron-rich Fe–O compounds at Earth's core pressures, The Innovation **4**, 100354 (2023).

[12] B. G. Jang, Y. He, J. H. Shim, H. Mao, and D. Y. Kim, Oxygen-Driven Enhancement of the Electron Correlation in Hexagonal Iron at Earth's Inner Core Conditions, J. Phys. Chem. Lett. **14**, 3884 (2023).

[13] G. L. Weerasinghe, C. J. Pickard, and R. J. Needs, Computational searches for iron oxides at high pressures, J. Phys.: Condens. Matter **27**, 455501 (2015).

[14] F. Zheng, Y. Sun, R. Wang, Y. Fang, F. Zhang, B. Da, S. Wu, C.-Z. Wang, R. M. Wentzcovitch, and K.-M. Ho, Structure and motifs of iron oxides from 1 to 3 TPa, Phys. Rev. Materials **6**, 043602 (2022).

[15] Y. Sun, M. Cococcioni, and R. M. Wentzcovitch, LDA + U s c calculations of phase relations in FeO, Phys. Rev. Materials **4**, 063605 (2020).

[16] Z. Zhang, Y. Sun, and R. M. Wentzcovitch, PBE-GGA predicts the B8↔B2 phase boundary of FeO at Earth's core conditions, Proc. Natl. Acad. Sci. U.S.A. **120**, e2304726120 (2023).

[17] T. Sun, J. P. Brodholt, Y. Li, and L. Vočadlo, Melting properties from *ab initio* free energy calculations: Iron at the Earth's inner-core boundary, Phys. Rev. B **98**, 224301 (2018).

[18] J. Zhuang, H. Wang, Q. Zhang, and R. M. Wentzcovitch, Thermodynamic properties of $\varepsilon$-Fe with thermal electronic excitation effects on vibrational spectra, Physical Review B **103**, 144102 (2021).

[19] Y. Sun, M. I. Mendelev, F. Zhang, X. Liu, B. Da, C. Wang, R. M. Wentzcovitch, and K. Ho, *Ab Initio* Melting Temperatures of Bcc and Hcp Iron Under the Earth's Inner Core Condition, Geophysical Research Letters **50**, e2022GL102447 (2023).

[20] S. Q. Wu, M. Ji, C. Z. Wang, M. C. Nguyen, X. Zhao, K. Umemoto, R. M. Wentzcovitch, and K. M. Ho, An adaptive genetic algorithm for crystal structure prediction, J. Phys.: Condens. Matter **26**, 035402 (2014).

[21] X. Zhao et al., Exploring the Structural Complexity of Intermetallic Compounds by an Adaptive Genetic Algorithm, Phys. Rev. Lett. **112**, 045502 (2014).

[22] D. M. Deaven and K. M. Ho, Molecular Geometry Optimization with a Genetic Algorithm, Phys. Rev. Lett. **75**, 288 (1995).

[23] P. E. Blöchl, Projector augmented-wave method, Phys. Rev. B **50**, 17953 (1994).

[24] G. Kresse and J. Furthmüller, Efficient iterative schemes for *ab initio* total-energy calculations using a plane-wave basis set, Phys. Rev. B **54**, 11169 (1996).

[25] G. Kresse and J. Furthmüller, Efficiency of ab-initio total energy calculations for metals and semiconductors using a plane-wave basis set, Computational Materials Science **6**, 15 (1996).

[26] J. P. Perdew, K. Burke, and M. Ernzerhof, Generalized Gradient Approximation Made Simple, Phys. Rev. Lett. **77**, 3865 (1996).

[27] N. D. Mermin, Thermal Properties of the Inhomogeneous Electron Gas, Phys. Rev. **137**, A1441 (1965).

[28] R. M. Wentzcovitch, J. L. Martins, and P. B. Allen, Energy versus free-energy conservation in first-principles molecular dynamics, Phys. Rev. B **45**, 11372 (1992).

[29] A. Togo, First-principles Phonon Calculations with Phonopy and Phono3py, J. Phys. Soc. Jpn. **92**, 012001 (2023).

[30] A. Togo, L. Chaput, T. Tadano, and I. Tanaka, Implementation strategies in phonopy and phono3py, J. Phys.: Condens. Matter **35**, 353001 (2023).

[31] T. Sun, D.-B. Zhang, and R. M. Wentzcovitch, Dynamic stabilization of cubic Ca Si O 3 perovskite at high temperatures and pressures from *ab initio* molecular dynamics, Phys. Rev. B **89**, 094109 (2014).

[32] D.-B. Zhang, T. Sun, and R. M. Wentzcovitch, Phonon Quasiparticles and Anharmonic Free Energy in Complex Systems, Phys. Rev. Lett. **112**, 058501 (2014).

[33] Z. Zhang, D.-B. Zhang, T. Sun, and R. M. Wentzcovitch, phq: A Fortran code to compute phonon quasiparticle properties and dispersions, Computer Physics Communications **243**, 110 (2019).

[34] J.-W. Xian, J. Yan, H.-F. Liu, T. Sun, G.-M. Zhang, X.-Y. Gao, and H.-F. Song, Effect of anharmonicity on the hcp to bcc transition in beryllium at high-pressure and high-temperature conditions, Phys. Rev. B **99**, 064102 (2019).

[35] W. G. Hoover, Canonical dynamics: Equilibrium phase-space distributions, Phys. Rev. A **31**, 1695 (1985).

[36] S. Nosé, A unified formulation of the constant temperature molecular dynamics methods, The Journal of Chemical Physics **81**, 511 (1984).

[37] H. J. Monkhorst and J. D. Pack, Special points for Brillouin-zone integrations, Phys. Rev. B **13**, 5188 (1976).

[38] T. Sun, X. Shen, and P. B. Allen, Phonon quasiparticles and anharmonic perturbation theory tested by molecular dynamics on a model system, PHYSICAL REVIEW B (2010).

[39] Z. Zhang and R. M. Wentzcovitch, *Ab initio* anharmonic thermodynamic properties of cubic Ca Si O 3 perovskite, Phys. Rev. B **103**, 104108 (2021).

[40] J. Zhuang, H. Wang, Q. Zhang, and R. M. Wentzcovitch, Thermodynamic properties of ɛ-Fe with thermal electronic excitation effects on vibrational spectra, PHYSICAL REVIEW B (2021).

[41] Y.-Y. Ye, Y. Chen, K.-M. Ho, B. N. Harmon, and P.-A. Lindgrd, Phonon-phonon coupling and the stability of the high-temperature bcc phase of Zr, Phys. Rev. Lett. **58**, 1769 (1987).

[42] O. Hellman, I. A. Abrikosov, and S. I. Simak, Lattice dynamics of anharmonic solids from first principles, Phys. Rev. B **84**, 180301 (2011).

[43] J. D. Althoff, P. B. Allen, R. M. Wentzcovitch, and J. A. Moriarty, Phase diagram and thermodynamic properties of solid magnesium in the quasiharmonic approximation, Phys. Rev. B **48**, 13253 (1993).

[44] Z. Li and S. Scandolo, Competing Phases of Iron at Earth's Core Conditions From Deep-Learning-Aided *ab-initio* Simulations, Geophysical Research Letters **51**, e2024GL110357 (2024).

[45] F. González-Cataldo and B. Militzer, *Ab initio* determination of iron melting at terapascal pressures and Super-Earths core crystallization, Phys. Rev. Research **5**, 033194 (2023).

[46] G. L. Weerasinghe, C. J. Pickard, and R. J. Needs, Computational searches for iron oxides at high pressures, Journal of Physics: Condensed Matter **27**, 455501 (2015).

[47] F. Birch, Elasticity and constitution of the Earth's interior, J. Geophys. Res. **57**, 227 (1952).